\documentclass{IEEEcsmag}

\usepackage[bookmarksopen,bookmarksdepth=2,breaklinks=true,colorlinks,urlcolor=blue,linkcolor=blue,citecolor=blue]{hyperref}
\usepackage{xurl}
\usepackage{amsmath,amssymb,amsthm}

\usepackage{xparse}
\usepackage{upmath}
\usepackage[italic]{mathastext}
\usepackage{stfloats}
\usepackage{comment}
\usepackage{soul}
\usepackage[dvipsnames]{xcolor}
\usepackage[resetlabels,labeled]{multibib}
\newcites{T}{REFERENCES}
\usepackage{setspace}
\usepackage{ragged2e}

\usepackage[colorinlistoftodos, textwidth=\marginparwidth]{todonotes}

\let\labelindent\relax
\usepackage{enumitem}
\usepackage{paracol}
\usepackage[many]{tcolorbox}
\usepackage{lipsum}

\newtcolorbox{NewBox}[1]{%
  floatplacement={#1}, width=0.625\textwidth,
  colframe=gray!10!black,colback=orange!10!white,boxrule=1pt,arc=.2em,boxsep=-1.6mm,
  }

\usepackage[pscoord]{eso-pic}

\newcommand{\placetextbox}[3]{
  \setbox0=\hbox{#3}
  \AddToShipoutPictureFG*{
    \put(\LenToUnit{#1\paperwidth},\LenToUnit{#2\paperheight}){\vtop{{\null}\makebox[0pt][c]{#3}}}%
  }%
}%

\newcommand\rev[1]{{{#1}}} 

\newcommand{\ao}[1]{{\textcolor{black}{#1}}}
\newcommand{\aov}[1]{{\textcolor{black}{#1}}}
\newcommand{\aol}[1]{{\textcolor{black}{#1}}}
\newcommand{\sg}[1]{{\textcolor{black}{#1}}}
\newcommand{\ch}[1]{{\textcolor{black}{#1}}}

\newcommand{\affilCMU}{$^{\Join}$}

\newcommand{\affilBilkent}{$^{\triangledown}$}

\newcommand{\affilETH}{$^{\diamond}$}

\newcommand{\affilUIUC}{$^{\star}$} 

\jvol{40}
\jnum{5}
\paper{8}
\jmonth{March}
\jname{IEEE MICRO}
\pubyear{2020}

\usepackage{floatrow}
\begin{document}

\placetextbox{0.55}{0.87}{\textsf{\emph{This is an extended and updated version of a paper published in}}}%
\placetextbox{0.55}{0.85}{\textsf{\emph{IEEE Micro, vol. 40, no. 5, pp. 65-75, 1 Sept.-Oct. 2020}}}%
\placetextbox{0.55}{0.83}{\textsf{\emph{\url{https://doi.org/10.1109/MM.2020.3013728}}}}%

\title{
\vspace{40pt}
\begin{center}\textbf{
\scalebox{2}{\fontsize{13pt}{0pt}\selectfont \textbf{Accelerating Genome Analysis:}}
\scalebox{2}{\fontsize{13pt}{0pt}\selectfont \textbf{A Primer on an Ongoing Journey}}
} 
\end{center}}

\author{
\vspace{-20pt}
\large
\begin{center}
{Mohammed Alser\affilETH}\quad%
{Z{\"u}lal Bing{\"o}l\affilBilkent}\quad%
{Damla Senol Cali\affilCMU}\quad%
{Jeremie Kim\affilETH\affilCMU}\quad\\
\vspace{5pt}
{Saugata Ghose\affilUIUC\affilCMU}\quad%
{Can Alkan\affilBilkent}\quad%
{Onur Mutlu\affilETH\affilCMU\affilBilkent}\\%
\end{center}
}
\affil{\normalsize
\begin{center}
\it\affilETH ETH Z{\"u}rich \quad 
\affilBilkent Bilkent University   \quad
\affilCMU Carnegie Mellon University\\
\vspace{2pt}
\quad \affilUIUC University of Illinois at Urbana–Champaign
\end{center}
}

\markboth{Mohammed Alser \ch{et al.}}{Accelerating Genome Analysis: A Primer on an Ongoing Journey}

\begin{abstract}
\justifying
\sg{Genome analysis} fundamentally starts with a process known as \emph{read mapping}, where \sg{sequenced fragments of an organism's genome are} 
compared against a reference \sg{genome.} 
Read mapping is currently a major bottleneck in the entire genome analysis pipeline, because state-of-the-art genome sequencing technologies are able to sequence a genome much faster than the computational techniques employed to analyze the genome. 
We describe the ongoing journey in significantly improving the performance of read mapping. 
We explain state-of-the-art algorithmic methods and hardware-based acceleration approaches. Algorithmic approaches exploit the structure of the genome as well as the structure of the underlying hardware.
Hardware-based acceleration approaches exploit specialized microarchitectures or various execution paradigms (e.g., processing inside or near memory). 
We conclude with the challenges of adopting these hardware-accelerated read mappers.
\end{abstract}

\maketitle

\chapterinitial{Genome analysis} is the foundation of many scientific and medical discoveries, \sg{and serves as a key enabler} of personalized medicine. This analysis is currently limited by the inability of modern genome sequencing technologies to read an organism's complete genome. Instead, sequencing machines extract smaller random fragments of an organism's DNA sequence, known as \emph{reads}. While the human genome contains over three billion \emph{bases} (i.e., A, C, G, T in DNA), the length of a read is orders of magnitude smaller, ranging from a few hundred bases (for \emph{short} reads) to a few million bases (for \emph{long} reads). Computers are used to \sg{perform genome \emph{assembly}, which reassembles} read fragments back into an entire genome sequence. 
\sg{Genome assembly} is currently the bottleneck to quickly and accurately determining an individual's entire genome, \sg{due to the complex algorithms and large datasets used for assembly}.


A widely-used approach for genome assembly is to perform \emph{sequence alignment}, which compares read fragments against a \sg{known} \textit{reference genome} (i.e., a complete representative DNA sequence for a particular species\rev{)}.  \sg{A process known as \emph{read mapping} matches each read generated from sequencing to one or more possible locations within the reference genome,}
based on the similarity between the read and the reference sequence segment at that location.
Unfortunately, the bases in a read may not be identical to the bases in the reference genome at the location that the read actually comes from.  These differences may be due to (1)~sequencing errors (up to 0.1\% in short reads~\cite{glenn2011field} and up to 20\% in long reads~\cite{ardui2018single, jain2018nanopore})
during extraction, and (2)~genetic mutations that are specific to the individual organism's DNA and may not exist in the reference genome.
Due to these \sg{potential} differences, the similarity between a read and a reference sequence segment must be identified using an \emph{approximate string matching} \sg{(ASM)} algorithm. 
\rev{The possible genetic differences between the reference genome and the sequenced genome are then identified using \emph{genomic variant calling} algorithms~\cite{ho2019structural}.}

\sg{The ASM performed during read mapping typically}
uses \sg{a computationally-expensive} 
\rev{dynamic programming (DP)} algorithm. 
This time-consuming algorithm \ch{has long been} a major bottleneck in the \ao{\emph{entire}} genome analysis pipeline, accounting for over 70\% of the execution time of read mapping~\cite{alser2019shouji}.
\rev{The vast majority of read mappers, such as the widely-used minimap2~\mbox{\cite{li2018minimap2}}, are implemented as software running on CPUs.}
\rev{We refer readers to a comprehensive survey~\mbox{\cite{alser2020technology}} for a discussion of state-of-the-art CPU-based read mappers.}
\sg{Accelerating ASM can help bridge the wide performance gap between sequencing machines and CPU-based read mapping algorithms, but faces four key challenges:}

\begin{enumerate}
    \item
        \sg{Due to the large datasets that \ao{a read mapper operates on, it generates} a large amount of \emph{data movement} between the \ao{CPU} and main memory.
        The CPU accesses off-chip main memory through a pin-limited bus known as the
        \emph{memory channel}, and a high amount of data movement across the memory channel
        is}
        extremely costly in terms of \sg{both} execution time and energy~\ao{\cite{mutlu2019processing, ghose2019processing}}. 
        %
        
    \item
        Modern sequencing machines generate read fragments at an exponentially higher rate than prior sequencing technologies, with their growth far outpacing the growth in computational power in recent years~\rev{\mbox{\cite{stephens2015big}}}.
        \rev{For example, the Illumina NovaSeq 6000 system can sequence about 48 human whole genomes at $30\times$ genome coverage (the average number of times a genomic base is sequenced) in about two days.
        However, analyzing (performing mapping and variant calling) the sequencing data of a \ao{\emph{single}} \aov{human} genome requires over 32 CPU hours on a 48-core Intel Xeon processor, 23 of which are spent on read mapping~\mbox{\cite{goyal2017ultra}}}.
    
    \item
        The first two \sg{challenges worsen} when a \emph{metagenomic} sample is profiled, where the sample donor is unknown. This requires matching the extracted reads to \emph{thousands} of reference genomes~\cite{lapierre2020metalign}. 
    \item 
        There is also an urgent need for rapidly incorporating clinical DNA sequencing and analysis into clinical practice for rapid surveillance of disease outbreaks (e.g., COVID-19~\cite{bloom2020swab}) and early diagnosis of genetic disorders in critically ill infants~\cite{friedman2019genome}.
     
\end{enumerate}

Increasing the number of CPUs used for genome analysis decreases the overall analysis time, but significantly \mbox{\ch{increases}} energy consumption and hardware costs. Cloud computing platforms are a potential alternative to distribute the workload at a reasonable cost, but are disallowed \mbox{\ch{due to}} data protection guidelines in many countries~\cite{langmead2018cloud}.

As a result, there is a dire need for new computational techniques that can \ch{quickly} process and analyze \ch{a} tremendous \ch{number} of extracted reads in order to drive cutting-edge advances in the genetic applications space~\cite{turakhia2018darwin}.
Many \sg{works} boost the performance of existing and new read mappers using new algorithms, hardware/software co-design, and hardware \ao{accelerators}.
Our \emph{goal} \sg{in this work} is to \ao{survey a prominent set of these three types of acceleration efforts} for guiding the design of new \sg{highly-efficient} read mappers.
\emph{To this end}, we (1)~discuss various state-of-the-art mechanisms and techniques that improve the execution time of read mapping using different modern high-performance computing architectures, and
(2)~highlight the challenges, \ao{in the last section}, that system architects and programmers must address to enable the widespread adoption of hardware-accelerated read mappers.

\section{Read Mapping}

\begin{figure*}[b!]
\centerline{\includegraphics[scale=0.6]{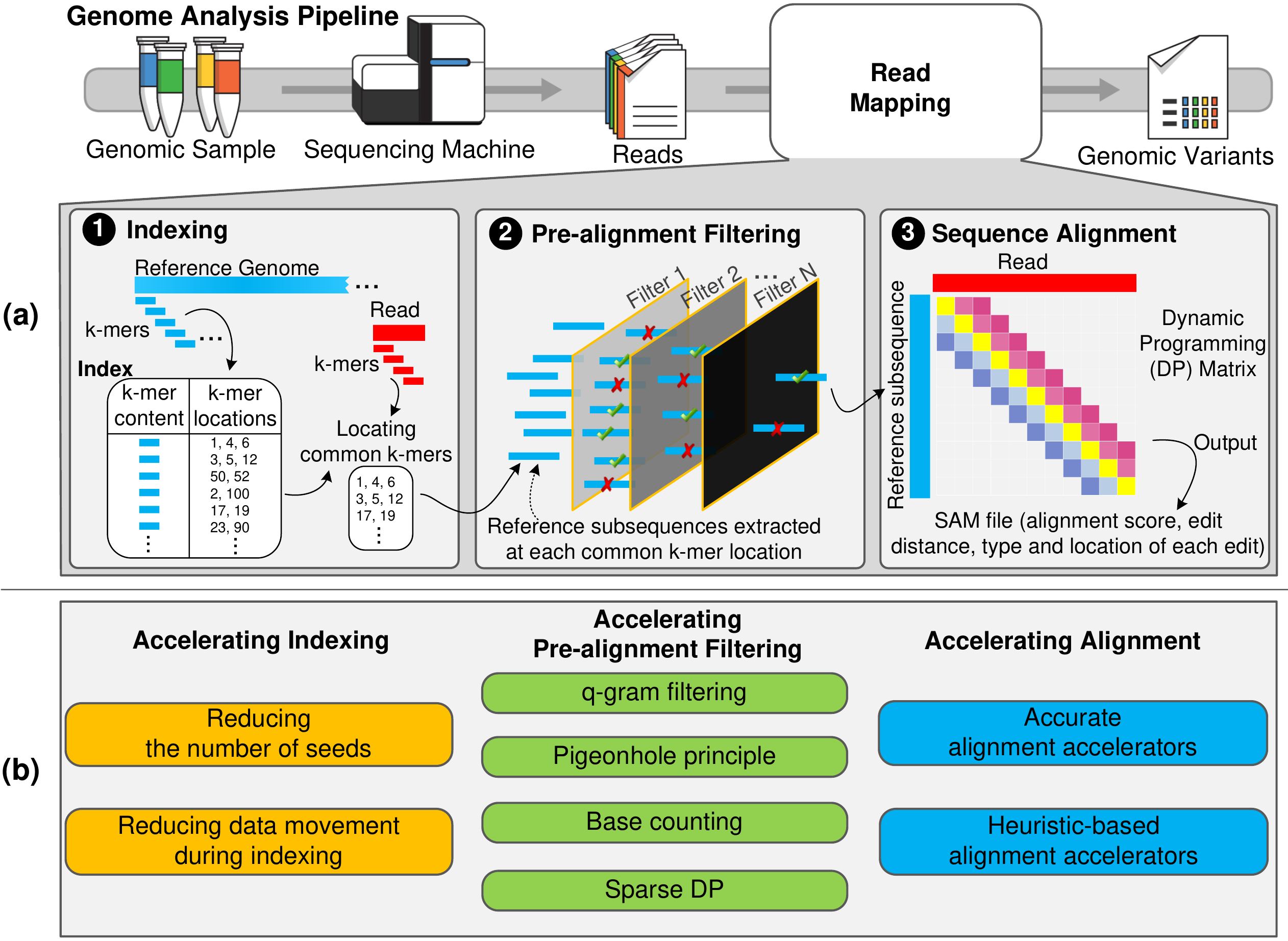}}

\caption{(a) The three steps of read mapping in genome analysis\ao{: (1) indexing, (2) pre-alignment filtering, and (3) sequence alignment}.
(b) \ao{Overview} of the existing approaches \aov{to} accelerating each step of read mapping.
}
\label{fig:steps}
\end{figure*}

The main goal of read mapping is to locate possible subsequences of the reference genome sequence that are similar to the read sequence while allowing at most $E$ edits\rev{, where $E$ is the \mbox{\ch{\emph{edit distance threshold}}}}.
Commonly allowed edits include deletion, insertion, and substitution of characters in one or both sequences. 
\vspace{50pt}

Edits can be as short as a single base pair (bp) alteration~\cite{nielsen2011genotype} or a much longer alteration (e.g., an insertion of about 600,000-base long region~\cite{jacquemont2011mirror}).
Mapping billions of reads to the reference genome \ch{is} computationally expensive~\ao{\cite{turakhia2018darwin, alser2019shouji,senolcali.micro2020}}.
Therefore, \sg{most read} mapping algorithms apply two key heuristic steps, \emph{indexing} and \emph{filtering}, to reduce the number of \sg{reference genome segments that need to}
be compared with each read.
\sg{The three steps of read mapping are shown in Figure~\ref{fig:steps}a.}
First, a \sg{read mapper indexes the reference genome by} using substrings (called \textit{seeds}) from each read to 
\sg{quickly identify all} potential mapping locations of each read in the reference genome.
Second, \sg{the mapper} uses filtering heuristics to examine the similarity \rev{for every sequence pair (a read sequence and one potential matching segment \mbox{\ch{in}} the reference genome identified during indexing)}. These filtering heuristics aim to eliminate most of the dissimilar sequence \rev{pairs}.
Third, \sg{the mapper performs sequence alignment (using ASM)} to check \ao{whether} \aov{or not} the remaining \rev{sequence} pairs that are identified by \sg{filtering to be similar} \aov{are} \ao{actually similar}.
\ao{The alignment step} examines all possible \emph{prefixes} of two sequences and tracks the prefixes that provide the highest possible \emph{alignment score} (known as \emph{optimal} alignment). 
The alignment score \sg{is a quantitative representation of the quality of an alignment for a given user-defined scoring function (computed based on the number of edits and/or matches)}.

Alignment algorithms typically use DP-based \ch{approaches} to avoid re-examining the same prefixes many times.
These DP-based algorithms provide the most accurate alignment results compared to other non-DP algorithms (such as the algorithm used in HISAT2~\cite{kim2019graph}), but they have quadratic time and space complexity (i.e., $O(m^2)$ for a sequence length of $m$).
Sequence alignment calculates information about the alignment such as the alignment score, \emph{edit distance}, and the type of each edit.
Edit distance is defined as the minimum number of changes needed to convert a sequence into the other sequence~\cite{levenshtein1966binary}.
\rev{Such} information is typically \sg{output by read mapping into a} \emph{sequence alignment/map} (SAM) file. 
\sg{Given the time spent on read mapping, all three steps have been targeted for acceleration.  Figure~\ref{fig:steps}b summarizes the different acceleration approaches, and we discuss a set of such works in the following sections.}

\section{Accelerating Indexing}
\sg{The indexing operation generates a table that is indexed by the contents of a seed, and identifies all locations where the seed exists in the reference genome.  Indexing needs to be done only once for a reference genome, and eliminates the need to perform ASM across the entire genome.}
\sg{During read mapping, a seed from a read is looked up in the table, and only \ao{the corresponding} locations are used for ASM (as only they can match the entire read).}
\rev{The major challenge with indexing is choosing the appropriate length and number of to-be-indexed seeds, as they can significantly impact the memory footprint and overall performance of read mapping~\mbox{\cite{li2018minimap2}}}. 
\rev{Querying short seeds potentially leads to a large number of mapping locations that need to be checked for a string match. 
The use of long reads requires extracting from each read \ao{a} large number of seeds, as the sequencing error rate is much higher in long reads.
This affects (1)~the number of times we query the index structure and (2)~the number of retrieved mapping locations.}
Thus, there are two key approaches used for accelerating the indexing step \ao{(Figure 1b)}. 


\subsection{{Reducing the Number of Seeds}}
\rev{Read mapping algorithms (e.g., minimap2~\mbox{\cite{li2018minimap2}}) typically reduce the number of seeds that are stored in the index structure by} finding the minimum representative set of seeds (called \emph{minimizers}) from a group of adjacent seeds within a genomic region.
\aov{The representative set can be calculated by imposing an ordering (e.g. lexicographically or by hash value) on a group of adjacent seeds and storing only the seed with the smallest order.}
\rev{Read mappers also apply heuristics to avoid} examining the mapping locations of a seed that \sg{occur more times} than a user-defined threshold value~\cite{li2018minimap2}.
Various data structures have been proposed and implemented  
to both reduce the storage cost of the indexing data structure and improve the algorithmic runtime of identifying the mapping locations within the indexing data structure.
\aov{One example of such data structures is FM-index (implemented by Langarita et al.~\cite{langarita2020compressed})\aol{, which} provides a compressed representation of the full-text index, while allowing for querying the index without the need for decompression.
This approach has two main advantages. 1) We can query seeds of arbitrary lengths, which helps to reduce the number of queried seeds. 2) It typically has less (by $1.5-2\times$) memory footprint compared to that of the indexing step of minimap2~\cite{li2018minimap2}.
However, one major bottleneck of FM-indexes is that locating the exact matches by querying the FM-index is significantly slower than that of classical indexes~\cite{vasimuddin2019efficient, langarita2020compressed}.
BWA-MEM2~\cite{vasimuddin2019efficient} proposes an uncompressed version of the FM-index that is at least $10\times$ larger than the compressed FM-index to speed up the querying step by $2\times$.
}
\subsection{{Reducing Data Movement During Indexing}}
RADAR~\cite{huangfu2018radar} observes that the indexing step is \sg{memory intensive, because the large number of random memory accesses dominates computation}. \sg{The} authors propose a processing-in-memory (PIM) architecture \ao{that} store\ao{s} the entire index inside the memory and enable\ao{s} querying the same index concurrently \ao{using} a large number of ASIC compute units. 
The amount of data movement is reduced from tens of gigabytes to a few bytes for a single query task, \sg{allowing RADAR to balance memory accesses with computation, and thus provide speedups and energy savings}.

\section{Accelerating Pre-Alignment Filtering}
After finding one or more \ao{potential} mapping locations of the read in the reference genome, \sg{the read mapper checks} the similarity between each read and each segment extracted at these mapping locations in the reference genome.
These segments can be \emph{similar} or \emph{dissimilar} to the read, though they share common seeds. 
To avoid examining dissimilar sequences using computationally-expensive sequence alignment algorithms, read mappers typically use filtering heuristics that are called \emph{pre-alignment filters}.

\ao{The key idea of pre-alignment filtering is to quickly estimate the number of edits between two given sequences and use this estimation to decide whether or not the computationally-expensive DP\aov{-based alignment} calculation is needed --- if not, a significant amount of time is saved by avoiding DP\aov{-based alignment}. 
If two genomic sequences differ by more than the edit distance threshold, then the two sequences are identified as dissimilar sequences and hence DP calculation is not needed}. 
In practice, only genomic sequence pairs with an edit distance less than or \rev{equal to} a user-defined threshold \ch{(i.e., \emph{E})} provide useful data for most \ao{genomic studies}~\cite{kim2018grim,alser2019shouji,alser2020technology}. 
\rev{Pre-alignment filters use one of four \ao{major} approaches to quickly filter out the dissimilar sequence pairs: (1)~the pigeonhole principle, (2)~base counting, (3)~$q$-gram filtering, or (4)~sparse DP.
Long read mappers typically use $q$-gram filtering or sparse DP, as their performance \ao{scales linearly with} read length and independent\aov{ly} of the edit distance}.


\subsection{Pigeonhole Principle}
The pigeonhole principle states that if $E$ items are put into $E$+1 boxes, then one or more boxes would be empty. 
\sg{This principle can be applied to \ao{detect dissimilar sequences and discard them from the candidate sequence pairs used for ASM}. If} two sequences differ by $E$ edits, then they should share at least a single subsequence (free of edits) among $E$+1 non-overlapping subsequences~\cite{alser2019shouji}, where $E$ is the \sg{edit distance threshold.}
\sg{For a read of length $m$}, if there are no more than $E$ edits between the read and the reference segment, then \sg{the read and reference segment are considered similar if they share} at most $E$+1 non-overlapping subsequences, with a total length of at least $m-E$.

\ch{The problem of identifying these $E$+1 non-overlapping subsequences is highly parallelizable, as these subsequences are independent of each other.
Shouji~\cite{alser2019shouji} exploits the pigeonhole principle to reduce the search space and provide a scalable architecture that can be implemented for any values of $m$ and $E$, by examining common subsequences independently and rapidly with high parallelism.
Shouji accelerates sequence alignment by 4.2-18.8$\times$ without affecting the alignment accuracy.}
\ch{We refer the reader to the sidebar for a brief discussion of \ao{several} other related works}.

\begin{NewBox}{s O{!htbp}}
\vspace{2pt}
\subsection{\textbf{\ch{Sidebar:} Related Works on Pre-Alignment Filtering Using the Pigeonhole Principle}}
\vspace{5pt}
\sg{Pigeonhole-filtering-based pre-alignment filtering can accelerate read mappers even without specialized hardware.  For example, the Adjacency Filter~[\textcolor{blue}{1}] accelerates} sequence alignment by up to 19$\times$. 
The accuracy and speed of pre-alignment filtering \sg{with} the pigeonhole principle have been rapidly improved over the last \ao{seven} years.
\sg{Shifted Hamming Distance (SHD)}~[\textcolor{blue}{2}] 
\sg{uses} SIMD-capable CPUs to provide high filtering speed, but \ao{supports a sequence length up to only 128 base pairs} due to the SIMD register \sg{width\ao{s}.}
GateKeeper~[\textcolor{blue}{3}] 
utilizes the large amounts of parallelism offered by FPGA architectures to accelerate SHD and \sg{overcome \ao{such} sequence length limitations.}
MAGNET~[\textcolor{blue}{4}] 
provides a comprehensive analysis of all sources of filtering inaccuracy of GateKeeper and SHD. 
Shouji~[\textcolor{blue}{5}] 
leverages this analysis to improve the accuracy of pre-alignment filtering by up to two orders of magnitude compared to both GateKeeper and SHD, using \sg{a} new algorithm and \ao{a} new FPGA architecture.
SneakySnake~[\textcolor{blue}{6}]
achieves up to four orders of magnitude higher filtering accuracy compared to GateKeeper and SHD by mapping the pre-alignment filtering problem to the single net routing (SNR) problem in VLSI chip layout. \sg{SNR finds} the shortest routing path that interconnects two terminals on the boundaries of \sg{a} VLSI chip layout \ao{in} the presence of obstacles. 
SneakySnake is the only pre-alignment filter that works on CPUs, GPUs, and FPGAs.
GenCache~[\textcolor{blue}{7}] 
proposes to perform highly-parallel pre-alignment filtering inside \sg{the CPU} cache to reduce data movement and improve energy efficiency, with about 20\% cache area overhead.
GenCache shows that using different existing pre-alignment filters together (similar approach to~\cite{rizk2010gassst}),
each of which operates only for a given edit distance threshold (e.g., \ao{using} SHD only when $E$ is between 1 and 5)\ao{,} provides a $2.5\times$ speedup over GenCache with a single pre-alignment filter.

\vspace{10pt}
{\fontfamily{phv}\selectfont
\large{$\blacksquare$} REFERENCES
\footnotesize{
\begin{enumerate}[label={\arabic*.},leftmargin=0.5cm]
\item Hongyi Xin et al. Accelerating Read Mapping with FastHASH. \textit{BMC Genomics}, 2013.
\item Hongyi Xin et al. Shifted Hamming Distance: A Fast and Accurate SIMD-Friendly Filter to Accelerate Alignment Verification in Read Mapping. \textit{Bioinformatics}, 2015.
\item Mohammed Alser et al. GateKeeper: A New Hardware Architecture for Accelerating Pre-Alignment in DNA Short Read Mapping. \textit{Bioinformatics}, 2017.
\item Mohammed Alser et al. MAGNET: Understanding and Improving the Accuracy of Genome Pre-Alignment Filtering. \textit{Transactions on Internet Research}, 2017.
\item Mohammed Alser et al. Shouji: A Fast and Efficient Pre-Alignment Filter for Sequence Alignment. \textit{Bioinformatics}, 2019.
\item Mohammed Alser et al. SneakySnake: A Fast and Accurate Universal Genome Pre-Alignment Filter for CPUs, GPUs, and FPGAs. arXiv:1910.09020 [q-bio.GN], 2019.
\item Anirban Nag et al. GenCache: Leveraging In-Cache Operators for Efficient Sequence Alignment. In \textit{MICRO}, 2019.
\end{enumerate}
}
}
\end{NewBox}

\subsection{Base Counting}
The base counting filter compares the numbers of \sg{bases (A, C, G, T)} in the read with \sg{\ao{the corresponding} base counts} in the reference segment. 
If one sequence has, for example, three more Ts than another sequence, then their alignment has at least three edits. 
If \sg{the difference in count is greater} than $E$, then the two sequences are dissimilar \sg{and the reference segment is discarded.}
The base counting filter is used in mrsFAST-Ultra~\cite{hach2014mrsfast} and GASSST~\mbox{\cite{rizk2010gassst}}.
\rev{Such a simple filtering approach rejects a significant fraction of dissimilar sequences (e.g., 49.8\%--80.4\% of sequences, as shown in GASSST~\mbox{\cite{rizk2010gassst}}) and thus avoids \ao{a large fraction of} expensive verification computations required by sequence alignment algorithms.}


\subsection{\emph{q}-gram Filtering Approach}
The $q$-gram filtering approach considers all of the sequence's possible overlapping substrings of length $q$ (known as $q$-grams).
Given a sequence of length $m$, there are $m-q+1$ overlapping $q$-grams that are obtained by sliding a window of length $q$ over the sequence. 
A single difference in one of the sequences can affect at most $q$ overlapping $q$-grams. 
Thus, $E$ differences can affect no more than $q\cdot E$ $q$-grams, where $E$ is the edit distance threshold. 
\sg{The} \emph{minimum number} of shared $q$-grams between two similar sequences \sg{is therefore} $(m-q+1)-(q\cdot E)$. 
This filtering approach requires very simple operations (e.g., sums and comparisons), which makes it attractive for hardware acceleration, such as in GRIM-Filter~\cite{kim2018grim}.
GRIM-Filter exploits the high memory bandwidth \sg{and \ao{computation capability in the} logic} layer of 3D-stacked memory to \sg{accelerate $q$-gram filtering} in the DRAM chip itself\ao{, using a new representation of reference genome that is friendly to in-memory processing}.
$q$-gram filtering is generally robust in handling only \sg{a} \ao{small} number of edits, as the presence of edits in any $q$-gram is significantly underestimated \ao{(e.g., counted as a single edit)}~\cite{weese2012razers}. 

\subsection{Sparse Dynamic Programming}
\sg{Sparse DP algorithms exploit} the exact matches (seeds) shared between a read and a reference segment to \sg{reduce execution time. These algorithms exclude} the corresponding locations of these seeds from estimating the number of edits between the two sequences,
\sg{as they were already detected as exact matches during indexing}.
\sg{Sparse} DP \ao{filtering techniques} apply DP-based alignment algorithms only between every two non-overlapping seeds to quickly estimate the total number of edits.
This approach is also known as \ao{\emph{chaining}}, and \sg{is} used in minimap2~\cite{li2018minimap2} and rHAT~\cite{liu2016rhat}. 
The recent work in~\cite{guo2019hardware} presents GPU 
and FPGA 
accelerators that achieve $7\times$ and $28\times$ acceleration, respectively, compared to the sequential implementation (executed with 14 CPU threads) of the chaining algorithm used in minimap2. 
 
\section{Accelerating Sequence Alignment}
After \sg{filtering out} most of the mapping locations that lead to dissimilar sequence pairs, \sg{read mapping calculates} the \ao{sequence} alignment information for every read and reference segment extracted at each mapping location.
\rev{Sequence alignment calculation is typically accelerated using one of two approaches: (1)~accelerating the DP-based algorithms using hardware accelerators without altering algorithm\mbox{\ch{ic}} behavior, and (2)~developing heuristics that sacrifice the optimality of the alignment score solution in order to reduce alignment time.

Despite more than three decades of attempts to accelerate sequence alignment, the fastest known edit distance algorithm~\cite{masek1980faster} has a nearly quadratic running time, $O(m^2/log_{2}m)$ for a sequence length of $m$, which is proven to be a tight bound~\cite{backurs2018edit}.
\aov{With maintaining the algorithm\mbox{\ch{ic}} behavior as in the first approach, it} \ao{is challenging} \aov{to} \ao{rapidly} \aov{calculate} \ao{sequence alignment of long reads with high parallelism.
\rev{As long reads have high sequencing error rates (up to 20\% of the read length), the edit distance threshold for long reads is typically higher than that for short reads, which results in calculating more entries in the DP matrix compared to that of short reads.}
The use of heuristics} \aov{(i.e., the second approach)} \ao{helps to reduce the number of calculated entries in the DP matrix and hence allows both the execution time and memory footprint to grow only linearly with read length} (as opposed to \ao{quadratically with classical} DP).}
Next, we describe \aov{the two} approaches in detail.

\subsection{{Accurate \ch{Alignment} Accelerators}}
From \sg{a} hardware perspective, sequence alignment \sg{acceleration has}  \rev{five} directions: (1)~using SIMD-capable CPUs, (2)~using multicore CPUs and GPUs, (3)~using FPGAs, \rev{(4)~using ASICs, and (5)~}using processing-in-memory architectures.
\sg{Traditional} DP-based algorithms are typically accelerated by computing only the necessary regions (i.e., diagonal vectors) of the DP matrix rather than the entire matrix, as proposed in Ukkonen's banded algorithm~\cite{ukkonen1985algorithms}.
This reduces the search space of the DP-based algorithm and \ao{reduces} computation time.
The number of diagonal bands required for computing the DP matrix is 2$E$+1, where $E$ is the edit distance threshold.
\rev{For example, the number of entries in the banded DP matrix for a 2~Mb long read can be 1.2 trillion.} 
Parasail~\cite{daily2016parasail} and KSW2 (used in minimap2~\cite{li2018minimap2}) exploit both Ukkonen's banded algorithm and SIMD-capable CPUs to compute \aov{banded} alignment for a sequence pair with a configurable scoring function. 
SIMD instructions offer significant parallelism to the matrix computation by executing the same vector operation on multiple operands at once. \rev{KSW2 is nearly as fast as Parasail when KSW2 does not use heuristics (explained in the next subsection)}.

\sg{The multicore} architecture of CPUs and GPUs provides the ability to compute alignments of many \sg{independent sequence pairs concurrently}. 
GASAL2~\cite{ahmed2019gasal2} exploits the multicore architecture of both CPUs and GPUs 
for highly-parallel computation of sequence alignment with a user-defined scoring function. 
Unlike other GPU-accelerated tools, \sg{GASAL2} transfers the bases to the GPU, without encoding them into binary format, and hides the data transfer time by overlapping \sg{GPU and CPU execution}. 
GASAL2 is up to $20\times$ faster than Parasail (\sg{when} executed with 56 CPU threads). 
\rev{BWA-MEM2~\mbox{\cite{vasimuddin2019efficient}} accelerates the banded sequence alignment \mbox{\ch{of its predecessor (BWA-MEM~\cite{li2013aligning})}} by up to $11.6\times$, by leveraging multicore and SIMD parallelism}.
\rev{However, to achieve \mbox{\ch{such levels of acceleration}}, BWA-MEM2 builds an index structure that is 6$\times$ larger than that of minimap2}.

Other designs, \sg{such as} FPGASW~\cite{fei2018fpgasw}, exploit the very large number of hardware execution units in FPGAs 
to form a linear systolic array~\cite{kung1982systolic}. 
Each execution unit in the systolic array is responsible for computing the value of a single entry of the DP matrix. The systolic array computes a single vector of the matrix at a time. The data dependency between the entries restricts the systolic array to computing the vectors sequentially (e.g., top-to-bottom, left-to-right, or in an anti-diagonal manner). 
FPGASW \sg{has a similar execution time} as its GPU implementation, 
but is $4\times$ more power efficient.

\rev{Specialized hardware accelerators (i.e., ASIC designs) provide application-specific, power- and area-efficient solutions to accelerate sequence alignment. For example, GenAx~\mbox{\cite{fujiki2018genax}}
is composed of SillaX, a sequence alignment accelerator, and a second accelerator for finding seeds. SillaX supports both a configurable scoring function and traceback operations.
SillaX is more efficient for short reads than for long reads, as it consists of an automata processor whose performance scales quadratically with the edit distance. 
GenAx is 31.7$\times$ faster than the predecessor of BWA-MEM2 (i.e., BWA-MEM~\cite{li2013aligning}) for short reads.}

\ao{Recent processing-in-memory architectures such as RAPID~\cite{gupta2019rapid} exploit the ability \aov{to perform} computation inside or near the memory chip to \aol{enable} efficient sequence alignment. 
RAPID modifies the DP-based alignment algorithm to make it friendly \aov{to} in-memory parallel computation by calculating two DP matrices\aov{: one} for calculating substitutions and exact matches and \aov{another} for calculating insertions and deletions. 
RAPID claims that this approach efficiently enables higher level\aov{s} of parallelism compared to traditional DP algorithms.
The main two \sg{benefits} of RAPID and such PIM-based architectures \aol{are higher performance and higher energy efficiency}~\cite{mutlu2019processing, ghose2019processing}, as they alleviate the need \sg{to transfer data between the main memory and} the CPU cores
through slow and energy hungry buses, while providing high degree of parallelism with the help of PIM. 
RAPID is on average $11.8\times$ faster and $212.7\times$ more power efficient than 384-GPU cluster of GPU implementation of sequence alignment, known as CUDAlign~\cite{de2016cudalign}.}

\subsection{{\ch{Heuristic-Based Alignment Accelerators}}}
The second direction is to \emph{limit} the functionality of the alignment algorithm or \emph{sacrifice} the optimality of the alignment solution \sg{in order to reduce} execution time.
\rev{The use of restrictive functionality and heuristics limits the possible applications of the algorithms that utilize this direction.}
\sg{Examples of limiting functionality include} limiting the scoring function, or only taking into account accelerating the computation of the DP matrix \rev{without performing the backtracking step~\mbox{\cite{chen2014accelerating}}}. 
There are several existing algorithms and \sg{corresponding} hardware accelerators that limit \sg{scoring function flexibility}. 
Levenshtein 
distance~\cite{levenshtein1966binary} 
and Myers's bit-vector algorithm 
\cite{myers1999fast} 
are examples of algorithms whose scoring functions are fixed, such that they penalize all types of edits equally when calculating the total alignment score. 
Restrictive scoring \sg{functions} reduce the total execution time of the alignment algorithm and reduce the bit-width requirement of the register that accommodates the value of each entry in the DP matrix. 
ASAP~\cite{banerjee2019asap} accelerates Levenshtein distance calculation by up to $63.3\times$ using FPGAs compared to its CPU implementation. 
The use of a fixed scoring function \ao{as in Edlib~\cite{vsovsic2017edlib}, which is the state-of-the-art implementation of Myers’s bit-vector algorithm,} helps to outperform Parasail (\sg{which uses a} flexible scoring function) by 12--1000$\times$.
\sg{One downside of fixed function scoring is that it may lead to the selection of a suboptimal sequence alignment.}

There are other algorithms and hardware architectures that provide low alignment time \sg{by trading off} accuracy.
Darwin~\cite{turakhia2018darwin} builds \ch{a} customized hardware architecture 
to speed up the alignment process, by dividing the DP matrix into overlapping submatrices and \sg{processing each submatrix} independently using systolic arrays.
Darwin provides three orders of magnitude speedup compared to Edlib~\cite{vsovsic2017edlib}.
Dividing the DP matrix (known as the \emph{Four-Russians Method}~\cite{arlazarov1970economical})
\sg{enables significant parallelism during} DP matrix computation, but it \ch{leads to} suboptimal alignment \rev{calculation~\mbox{\cite{rizk2010gassst}}}.
Darwin claims that choosing \sg{a} large submatrix size ($\geq 320\times320$) and ensuring sufficient overlap ($\geq$128 entries) between adjacent submatrices \ch{may} provide optimal alignment calculation for some datasets.

There are other proposals that limit the number of calculated entries of the DP matrix based on one of two approaches: (1)~using sparse DP or (2)~\ch{using a greedy approach to maintain} a high alignment score.
Both approaches suffer from providing suboptimal alignment \rev{calculation~\mbox{\cite{slater2005automated, zhang2000greedy}}}.
The first approach uses the same sparse DP algorithm used for pre-alignment filtering but as an alignment step, as \sg{done} in 
\ch{the {\tt exonerate} tool}~\cite{slater2005automated}.
The second approach is \ch{employed in} $X$-drop~\rev{\mbox{\cite{zhang2000greedy}}}, which (1)~avoids calculating entries (and their neighbors) whose alignment \sg{scores are} more than $X$ below the highest score seen so far \sg{(where $X$ is a user-specified parameter)}, and (2)~stops early when a high alignment score is not possible. 
The $X$-drop algorithm is guaranteed to find the optimal alignment between relatively-similar sequences for \rev{\emph{only some} scoring functions~\mbox{\cite{zhang2000greedy}}}. 
\rev{A similar algorithm (known as $Z$-drop) makes KSW2 at least} $2.6\times$ \rev{faster than Parasail}.
A recent GPU implementation~\cite{zeni2020logan} of the $X$-Drop algorithm is 3.1--120.4$\times$ faster than KSW2.

There are also a large number of edit distance approximation algorithms that provide a reduction in time complexity (e.g., $O(m^{1.647})$ instead of $O(m^{2})$), but they suffer from providing overestimated edit distance~\cite{batu2006oblivious,andoni2012approximating,chakraborty2018approximating,charikar2018estimating}.


\section{Discussion and Future Opportunities}
Despite more than two decades of attempts, bridging the performance gap between sequencing machines and read mapping is still challenging. \sg{We summarize four main challenges below}. 

First, we need to accelerate the \ao{entire read mapping process} rather than its individual steps.
Accelerating only a single step of read mapping \ch{limits} the overall achieved speedup according to Amdahl's Law.
\rev{Illumina and NVIDIA \ao{have} recently started follow\mbox{\ch{ing a more holistic}}}
\rev{approach, and they claim to accelerate genome analysis by more than $48\times$, mainly by using specialization and hardware/software co-design.}
\rev{Illumina has built an FPGA-based platform, called DRAGEN (}\url{https://www.illumina.com/products/by-type/informatics-products/dragen-bio-it-platform.html}\rev{)}\rev{, 
that accelerates all steps of genome analysis, including read mapping and variant calling. DRAGEN reduces the overall analysis time from 32 CPU hours 
to only 37 minutes~\mbox{\cite{goyal2017ultra}}}. 
\rev{NVIDIA has built Parabricks, a software suite accelerated using the company's latest GPUs.
Parabricks} ({\url{https://developer.nvidia.com/clara-parabricks}}\rev{) can analyze whole human genomes at 30$\times$ coverage in about 45 minutes}.

\ch{\rev{Second, we need to reduce the high amount of data movement that takes place during genome analysis. Moving data (1)~between compute units and main memory, (2)~between multiple hardware accelerators, and (3)~between the sequencing machine and the computer performing the analysis incurs high costs in terms of execution time and energy.
These costs are a significant barrier to enabling efficient analysis that can keep up with sequencing technologies, and some recent works try to tackle this problem \ao{\cite{kim2018grim, mutlu2019processing, ghose2019processing}}.
GenASM~\mbox{\cite{senolcali.micro2020}} is a framework that uses bitvector-based ASM to accelerate multiple steps of the genome analysis pipeline, and is designed to be implemented inside 3D-stacked memory. 
Through a combination of hardware--software co-design to unlock parallelism, and processing-in-memory to reduce data movement, GenASM can perform (1)~pre-alignment filtering for short reads, (2)~sequence alignment for both short and long reads, and (3)~whole genome alignment, among other use cases. 
For short/long read alignment, GenASM achieves 111$\times$/116$\times$ speedup over state-of-the-art software read mappers while reducing power consumption by 33$\times$/37$\times$.}}
\rev{DRAGEN reduces data movement between the sequencing machine and the computer performing analysis by adding specialized hardware support inside the sequencing machine for data compression.
However, this still requires movement of compressed data.
Performing read mapping inside the sequencing machine itself can significantly improve efficiency by eliminating sequencer-to-computer movement, and embedding a single specialized chip for read mapping within a portable sequencing device can potentially enable new applications of genome sequencing (e.g., rapid surveillance of diseases such as Ebola~\cite{quick2016real} and COVID-19~\cite{bloom2020swab}, near-patient testing, bringing precision medicine to remote locations).
Unfortunately, efforts in this direction remain very limited.}

\ch{Third, \rev{we need to develop flexible hardware architectures that do not conservatively limit the range of supported parameter values at design time.}}
\sg{Commonly}-used read mappers (e.g., minimap2) have different input parameters, each of which has a wide range of input values. 
For example, the edit distance threshold is typically user defined and can be very high (15-20\% of the read length) for recent long reads. 
\sg{A configurable} scoring function is another example, \sg{as it} determines the number of bits needed to store each entry of the DP matrix 
\rev{(e.g., DRAGEN imposes a restriction on the maximum frequency of seed occurrence)}.
\sg{Due to rapid} changes in sequencing technologies (e.g., high sequencing error rate and longer read lengths)~\cite{senol2019nanopore, firtina2020apollo},
\sg{these design restrictions can quickly make specialized hardware obsolete.}
Thus, read mappers need to adapt their algorithm\ch{s} and their hardware architecture\ch{s} \ao{to be modular and scalable so that they can be implemented for any sequence length and edit distance threshold based on the sequencing technology}.


Fourth, \rev{we need to adapt existing genomic data formats for hardware accelerators or develop more efficient file formats}. Most \sg{sequencing data is} stored in \sg{the} FASTQ/FASTA format, where each base takes a single byte (8 bits) of memory. 
\sg{This encoding is inefficient, as only 2~bits (3~bits when the ambiguous base, N, is included) are needed to encode each DNA base.}
The sequencing machine converts sequenced bases into \sg{FASTQ/FASTA format, and hardware accelerators convert the file contents into unique (for each accelerator) compact binary representations} for efficient processing.
This process \ch{that requires multiple format conversions} wastes \sg{time}.
\rev{For example, only 43\% of the sequence alignment time in BWA-MEM2~\cite{vasimuddin2019efficient} is spent on calculating the DP matrix, while 33\% of the sequence alignment time is spent on pre-processing the input sequences for loading into SIMD registers}\aov{, as provided in~\cite{vasimuddin2019efficient}}.
\rev{To address this inefficiency,} we need to widely adopt efficient \sg{hardware-friendly formats}, such as UCSC's 2bit \ao{format} \rev{(\mbox{\textcolor{blue}{https://genome.ucsc.edu/goldenPath/help/twoBit}})}\sg{, to maximize the benefits of hardware accelerators and reduce resource utilization.}
\rev{We are not aware of any recent read mapper that uses such formats.}

\sg{The acceleration efforts we highlight in this work represent state-of-the-art efforts to reduce current bottlenecks in the genome analysis pipeline.
We hope that these efforts and the challenges we discuss provide a foundation for future work in accelerating read mappers and developing other genome sequence analysis tools.}

\section{Acknowledgments}
The work of Onur Mutlu's SAFARI Research Group was supported by funding from Intel, the Semiconductor Research Corporation, VMware, and the National Institutes of Health (NIH).

\bibliographystyle{unsrt}
\bibliography{document}

\begin{thebibliography}{10}

\bibitem{glenn2011field}
Travis~C Glenn.
\newblock {Field Guide to Next-Generation DNA Sequencers}.
\newblock {\em Molecular Ecology Resources}, 2011.

\bibitem{ardui2018single}
Simon Ardui, Adam Ameur, Joris~R Vermeesch, and Matthew~S Hestand.
\newblock {Single Molecule Real-Time (SMRT) Sequencing Comes of Age:
  Applications and Utilities for Medical Diagnostics}.
\newblock {\em Nucleic Acids Research}, 2018.

\bibitem{jain2018nanopore}
Miten Jain, Sergey Koren, Karen~H Miga, Josh Quick, Arthur~C Rand, Thomas~A
  Sasani, John~R Tyson, Andrew~D Beggs, Alexander~T Dilthey, Ian~T Fiddes,
  et~al.
\newblock Nanopore sequencing and assembly of a human genome with ultra-long
  reads.
\newblock {\em Nature biotechnology}, 2018.

\bibitem{ho2019structural}
Steve~S Ho, Alexander~E Urban, and Ryan~E Mills.
\newblock {Structural variation in the sequencing era}.
\newblock {\em Nature Reviews Genetics}, pages 1--19, 2019.

\bibitem{alser2019shouji}
Mohammed Alser, Hasan Hassan, Akash Kumar, Onur Mutlu, and Can Alkan.
\newblock {Shouji: A Fast and Efficient Pre-Alignment Filter for Sequence
  Alignment}.
\newblock {\em Bioinformatics}, 2019.

\bibitem{li2018minimap2}
Heng Li.
\newblock {minimap2: Pairwise Alignment for Nucleotide Sequences}.
\newblock {\em Bioinformatics}, 2018.

\bibitem{alser2020technology}
Mohammed Alser, Jeremy Rotman, Kodi Taraszka, Huwenbo Shi, Pelin~Icer Baykal,
  Harry~Taegyun Yang, Victor Xue, Sergey Knyazev, Benjamin~D. Singer, Brunilda
  Balliu, David Koslicki, Pavel Skums, Alex Zelikovsky, Can Alkan, Onur Mutlu,
  and Serghei Mangul.
\newblock {Technology Dictates Algorithms: Recent Developments in Read
  Alignment}.
\newblock arXiv:2003.00110 [q-bio.GN], 2020.

\bibitem{mutlu2019processing}
Onur Mutlu, Saugata Ghose, Juan G{\'o}mez-Luna, and Rachata Ausavarungnirun.
\newblock Processing data where it makes sense: Enabling in-memory computation.
\newblock {\em Microprocessors and Microsystems}, 67:28--41, 2019.

\bibitem{ghose2019processing}
Saugata Ghose, Amirali Boroumand, Jeremie~S Kim, Juan G{\'o}mez-Luna, and Onur
  Mutlu.
\newblock {Processing-in-memory: A workload-driven perspective}.
\newblock {\em IBM Journal of Research and Development}, 63(6):3--1, 2019.

\bibitem{stephens2015big}
Zachary~D. Stephens, Skylar~Y. Lee, Faraz Faghri, Roy~H. Campbell, Chengxiang
  Zhai, Miles~J. Efron, Ravishankar Iyer, Michael~C. Schatz, Saurabh Sinha, and
  Gene~E. Robinson.
\newblock {Big Data: Astronomical or Genomical?}
\newblock {\em PLoS Biology}, 2015.

\bibitem{goyal2017ultra}
Amit Goyal, Hyuk~Jung Kwon, Kichan Lee, Reena Garg, Seon~Young Yun, Yoon~Hee
  Kim, Sunghoon Lee, and Min~Seob Lee.
\newblock {Ultra-Fast Next Generation Human Genome Sequencing Data Processing
  Using DRAGEN\textsuperscript{\textregistered} Bio-IT Processor for Precision
  Medicine}.
\newblock {\em Open Journal of Genetics}, 2017.

\bibitem{lapierre2020metalign}
Nathan LaPierre, Mohammed Alser, Eleazar Eskin, David Koslicki, and Serghei
  Mangul.
\newblock {Metalign: Efficient alignment-based metagenomic profiling via
  containment min hash}.
\newblock {\em BioRxiv}, 2020.

\bibitem{bloom2020swab}
Joshua~S Bloom, Eric~M Jones, Molly Gasperini, Nathan~B Lubock, Laila Sathe,
  Chetan Munugala, A~Sina Booeshaghi, Oliver~F Brandenberg, Longhua Guo, James
  Boocock, et~al.
\newblock Swab-seq: A high-throughput platform for massively scaled up
  sars-cov-2 testing.
\newblock {\em medRxiv}, 2020.

\bibitem{friedman2019genome}
Jan~M Friedman, Yvonne Bombard, Martina~C Cornel, Conrad~V Fernandez, Anne~K
  Junker, Sharon~E Plon, Zornitza Stark, and Bartha~Maria Knoppers.
\newblock {Genome-wide sequencing in acutely ill infants: genomic medicine’s
  critical application?}
\newblock {\em Genetics in Medicine}, 21(2):498--504, 2019.

\bibitem{langmead2018cloud}
Ben Langmead and Abhinav Nellore.
\newblock {Cloud Computing for Genomic Data Analysis and Collaboration}.
\newblock {\em Nature Reviews Genetics}, 2018.

\bibitem{turakhia2018darwin}
Yatish Turakhia, Gill Bejerano, and William~J. Dally.
\newblock {Darwin: A Genomics Co-Processor Provides Up to 15,000X Acceleration
  on Long Read Assembly}.
\newblock In {\em ASPLOS}, 2018.

\bibitem{nielsen2011genotype}
Rasmus Nielsen, Joshua~S Paul, Anders Albrechtsen, and Yun~S Song.
\newblock {Genotype and SNP calling from next-generation sequencing data}.
\newblock {\em Nature Reviews Genetics}, 12(6):443--451, 2011.

\bibitem{jacquemont2011mirror}
S{\'e}bastien Jacquemont, Alexandre Reymond, Flore Zufferey, Louise Harewood,
  Robin~G Walters, Zolt{\'a}n Kutalik, Danielle Martinet, Yiping Shen, Armand
  Valsesia, Noam~D Beckmann, et~al.
\newblock {Mirror extreme BMI phenotypes associated with gene dosage at the
  chromosome 16p11. 2 locus}.
\newblock {\em Nature}, 478(7367):97--102, 2011.

\bibitem{senolcali.micro2020}
Damla~Senol Cali, Gurpreet~S. Kalsi, Z{\"u}lal Bing{\"o}l, Lavanya Subramanian,
  Can Firtina, Jeremie~S. Kim, Rachata Ausavarungnirun, Mohammed Alser, Anant
  Nori, Juan~G{\'o}mez Luna, Amirali Boroumand, Allison Scibisz, Sreenivas
  Subramoney, Can Alkan, Saugata Ghose, and Onur Mutlu.
\newblock {GenASM: A Low-Power, Memory-Efficient Approximate String Matching
  Acceleration Framework for Genome Sequence Analysis}.
\newblock In {\em Proc. 53rd Int. Symp. Microarchitecture (MICRO)}, 2020.

\bibitem{kim2019graph}
Daehwan Kim, Joseph~M Paggi, Chanhee Park, Christopher Bennett, and Steven~L
  Salzberg.
\newblock {Graph-based genome alignment and genotyping with HISAT2 and
  HISAT-genotype}.
\newblock {\em Nature biotechnology}, 37(8):907--915, 2019.

\bibitem{levenshtein1966binary}
Vladimir~I Levenshtein.
\newblock {Binary codes capable of correcting deletions, insertions, and
  reversals}.
\newblock In {\em Soviet Physics-Doklady}, volume~10, pages 707--710, 1966.

\bibitem{langarita2020compressed}
Ruben Langarita, Adria Armejach, Javier Setoain, Pablo Enrique~Ibanez Marin,
  Jes{\'u}s Alastruey-Bened{\'e}, and Miquel~Moreto Planas.
\newblock {Compressed Sparse FM-Index: Fast Sequence Alignment Using Large
  K-Steps}.
\newblock {\em IEEE/ACM Transactions on Computational Biology and
  Bioinformatics}, 2020.

\bibitem{vasimuddin2019efficient}
Md~Vasimuddin, Sanchit Misra, Heng Li, and Srinivas Aluru.
\newblock {Efficient Architecture-Aware Acceleration of BWA-MEM for Multicore
  Systems}.
\newblock In {\em IPDPS}, 2019.

\bibitem{huangfu2018radar}
Wenqin Huangfu, Shuangchen Li, Xing Hu, and Yuan Xie.
\newblock {RADAR: A 3D-ReRAM Based DNA Alignment Accelerator Architecture}.
\newblock In {\em DAC}, 2018.

\bibitem{kim2018grim}
Jeremie~S. Kim, Damla Senol~Cali, Hongyi Xin, Donghyuk Lee, Saugata Ghose,
  Mohammed Alser, Hasan Hassan, Oguz Ergin, Can Alkan, and Onur Mutlu.
\newblock {GRIM-Filter: Fast Seed Location Filtering in DNA Read Mapping Using
  Processing-in-Memory Technologies}.
\newblock {\em BMC Genomics}, 2018.

\bibitem{rizk2010gassst}
Guillaume Rizk and Dominique Lavenier.
\newblock {GASSST: Global Alignment Short Sequence Search Tool}.
\newblock {\em Bioinformatics}, 2010.

\bibitem{hach2014mrsfast}
Faraz Hach, Iman Sarrafi, Farhad Hormozdiari, Can Alkan, Evan~E Eichler, and
  S~Cenk Sahinalp.
\newblock {mrsFAST-Ultra: a compact, SNP-aware mapper for high performance
  sequencing applications}.
\newblock {\em Nucleic acids research}, 42(W1):W494--W500, 2014.

\bibitem{weese2012razers}
David Weese, Manuel Holtgrewe, and Knut Reinert.
\newblock {RazerS 3: faster, fully sensitive read mapping}.
\newblock {\em Bioinformatics}, 28(20):2592--2599, 2012.

\bibitem{liu2016rhat}
Bo~Liu, Dengfeng Guan, Mingxiang Teng, and Yadong Wang.
\newblock {rHAT: fast alignment of noisy long reads with regional hashing}.
\newblock {\em Bioinformatics}, 32(11):1625--1631, 2016.

\bibitem{guo2019hardware}
Licheng Guo, Jason Lau, Zhenyuan Ruan, Peng Wei, and Jason Cong.
\newblock {Hardware acceleration of long read pairwise overlapping in genome
  sequencing: A race between FPGA and GPU}.
\newblock In {\em 2019 IEEE 27th Annual International Symposium on
  Field-Programmable Custom Computing Machines (FCCM)}, pages 127--135. IEEE,
  2019.

\bibitem{masek1980faster}
William~J Masek and Michael~S Paterson.
\newblock {A faster algorithm computing string edit distances}.
\newblock {\em Journal of Computer and System sciences}, 20(1):18--31, 1980.

\bibitem{backurs2018edit}
Arturs Backurs and Piotr Indyk.
\newblock Edit distance cannot be computed in strongly subquadratic time
  (unless {SETH} is false).
\newblock {\em SIAM Journal on Computing}, 47(3):1087--1097, 2018.

\bibitem{ukkonen1985algorithms}
Esko Ukkonen.
\newblock {Algorithms for approximate string matching}.
\newblock {\em Information and control}, 64(1-3):100--118, 1985.

\bibitem{daily2016parasail}
Jeff Daily.
\newblock {Parasail: SIMD C Library for Global, Semi-Global, and Local Pairwise
  Sequence Alignments}.
\newblock {\em BMC Bioinformatics}, 2016.

\bibitem{ahmed2019gasal2}
Nauman Ahmed, Jonathan L{\'e}vy, Shanshan Ren, Hamid Mushtaq, Koen Bertels, and
  Zaid Al-Ars.
\newblock {GASAL2: A GPU Accelerated Sequence Alignment lLibrary for
  High-Throughput NGS Data}.
\newblock {\em BMC Bioinformatics}, 2019.

\bibitem{li2013aligning}
Heng Li.
\newblock Aligning sequence reads, clone sequences and assembly contigs with
  bwa-mem.
\newblock {\em arXiv preprint arXiv:1303.3997}, 2013.

\bibitem{fei2018fpgasw}
Xia Fei, Zou Dan, Lu~Lina, Man Xin, and Zhang Chunlei.
\newblock {FPGASW: Accelerating Large-Scale Smith--Waterman Sequence Alignment
  Application with Backtracking on FPGA Linear Systolic Array}.
\newblock {\em Interdisciplinary Sciences: Computational Life Sciences}, 2018.

\bibitem{kung1982systolic}
HT~Kung.
\newblock {Why Systolic Architectures?}
\newblock {\em IEEE Computer}, (1):37--46, 1982.

\bibitem{fujiki2018genax}
Daichi Fujiki, Arun Subramaniyan, Tianjun Zhang, Yu~Zeng, Reetuparna Das, David
  Blaauw, and Satish Narayanasamy.
\newblock {GenAx: A Genome Sequencing Accelerator}.
\newblock In {\em ISCA}, 2018.

\bibitem{gupta2019rapid}
Saransh Gupta, Mohsen Imani, Behnam Khaleghi, Venkatesh Kumar, and Tajana
  Rosing.
\newblock {RAPID: A ReRAM Processing in-Memory Architecture for DNA Sequence
  Alignment}.
\newblock In {\em 2019 IEEE/ACM International Symposium on Low Power
  Electronics and Design (ISLPED)}, pages 1--6. IEEE, 2019.

\bibitem{de2016cudalign}
Edans~Flavius de~Oliveira~Sandes, Guillermo Miranda, Xavier Martorell, Eduard
  Ayguade, George Teodoro, and Alba Cristina~Magalhaes Melo.
\newblock {CUDAlign 4.0: Incremental speculative traceback for exact
  chromosome-wide alignment in GPU clusters}.
\newblock {\em IEEE Transactions on Parallel and Distributed Systems},
  27(10):2838--2850, 2016.

\bibitem{chen2014accelerating}
Peng Chen, Chao Wang, Xi~Li, and Xuehai Zhou.
\newblock {Accelerating the Next Generation Long Read Mapping with the
  FPGA-Based System}.
\newblock {\em IEEE/ACM Transactions on Computational Biology and
  Bioinformatics}, 2014.

\bibitem{myers1999fast}
Gene Myers.
\newblock {A fast bit-vector algorithm for approximate string matching based on
  dynamic programming}.
\newblock {\em Journal of the ACM (JACM)}, 46(3):395--415, 1999.

\bibitem{banerjee2019asap}
Subho~Sankar Banerjee, Mohamed El-Hadedy, Jong~Bin Lim, Zbigniew~T. Kalbarczyk,
  Deming Chen, Steven~S. Lumetta, and Ravishankar~K. Iyer.
\newblock {ASAP: Accelerated Short-Read Alignment on Programmable Hardware}.
\newblock {\em IEEE Transactions on Computers}, 2019.

\bibitem{vsovsic2017edlib}
Martin {\v{S}}o{\v{s}}i{\'c} and Mile {\v{S}}iki{\'c}.
\newblock {Edlib: A C/C++ Library for Fast, Exact Sequence Alignment Using Edit
  Distance}.
\newblock {\em Bioinformatics}, 2017.

\bibitem{arlazarov1970economical}
Vladimir~L'vovich Arlazarov, Yefim~A Dinitz, MA~Kronrod, and IgorAleksandrovich
  Faradzhev.
\newblock {On economical construction of the transitive closure of an oriented
  graph}.
\newblock In {\em Doklady Akademii Nauk}, volume 194, pages 487--488. Russian
  Academy of Sciences, 1970.

\bibitem{slater2005automated}
Guy St.~C. Slater and Ewan Birney.
\newblock {Automated Generation of Heuristics for Biological Sequence
  Comparison}.
\newblock {\em BMC Bioinformatics}, 2005.

\bibitem{zhang2000greedy}
Zheng Zhang, Scott Schwartz, Lukas Wagner, and Webb Miller.
\newblock {A Greedy Algorithm for Aligning DNA Sequences}.
\newblock {\em Journal of Computational Biology}, 2000.

\bibitem{zeni2020logan}
Alberto Zeni, Giulia Guidi, Marquita Ellis, Nan Ding, Marco~D. Santambrogio,
  Steven Hofmeyr, Ayd{\i}n Bulu{\c{c}}, Leonid Oliker, and Katherine Yelick.
\newblock {LOGAN: High-Performance GPU-Based X-Drop Long-Read Alignment}.
\newblock arXiv:2002.05200 [q-bio.GN], 2020.

\bibitem{batu2006oblivious}
Tu{\u{g}}kan Batu, Funda Ergun, and Cenk Sahinalp.
\newblock {Oblivious string embeddings and edit distance approximations}.
\newblock In {\em Proceedings of the seventeenth annual ACM-SIAM Symposium on
  Discrete Algorithms (SODA)}, pages 792--801. Society for Industrial and
  Applied Mathematics, 2006.

\bibitem{andoni2012approximating}
Alexandr Andoni and Krzysztof Onak.
\newblock {Approximating edit distance in near-linear time}.
\newblock {\em SIAM Journal on Computing}, 41(6):1635--1648, 2012.

\bibitem{chakraborty2018approximating}
Diptarka Chakraborty, Debarati Das, Elazar Goldenberg, Michal Koucky, and
  Michael Saks.
\newblock {Approximating edit distance within constant factor in truly
  sub-quadratic time}.
\newblock In {\em 2018 IEEE 59th Annual Symposium on Foundations of Computer
  Science (FOCS)}, pages 979--990. IEEE, 2018.

\bibitem{charikar2018estimating}
Moses Charikar, Ofir Geri, Michael~P Kim, and William Kuszmaul.
\newblock {On Estimating Edit Distance: Alignment, Dimension Reduction, and
  Embeddings}.
\newblock In {\em 45th International Colloquium on Automata, Languages, and
  Programming (ICALP 2018)}. Schloss Dagstuhl-Leibniz-Zentrum fuer Informatik,
  2018.

\bibitem{quick2016real}
Joshua Quick, Nicholas~J Loman, Sophie Duraffour, Jared~T Simpson, Ettore
  Severi, Lauren Cowley, Joseph~Akoi Bore, Raymond Koundouno, Gytis Dudas, Amy
  Mikhail, et~al.
\newblock {Real-time, portable genome sequencing for Ebola surveillance}.
\newblock {\em Nature}, 530(7589):228--232, 2016.

\bibitem{senol2019nanopore}
Damla Senol~Cali, Jeremie~S Kim, Saugata Ghose, Can Alkan, and Onur Mutlu.
\newblock {Nanopore sequencing technology and tools for genome assembly:
  Computational analysis of the current state, bottlenecks and future
  directions}.
\newblock {\em Briefings in Bioinformatics}, 20(4):1542--1559, 2019.

\bibitem{firtina2020apollo}
Can Firtina, Jeremie~S Kim, Mohammed Alser, Damla Senol~Cali, A~Ercument Cicek,
  Can Alkan, and Onur Mutlu.
\newblock {Apollo: a sequencing-technology-independent, scalable and accurate
  assembly polishing algorithm}.
\newblock {\em Bioinformatics}, 36(12):3669--3679, 2020.

\end{thebibliography}
\begin{IEEEbiography}{Mohammed Alser}{\,}is with ETH Zürich, Switzerland. Contact him at alserm@inf.ethz.ch
\end{IEEEbiography}

\begin{IEEEbiography}{Zülal Bingöl}{\,}is with Bilkent University, Turkey. Contact her at zulal.bingol@bilkent.edu.tr
\end{IEEEbiography}

\begin{IEEEbiography}{Damla Senol Cali}{\,}is with Carnegie Mellon University, USA. Contact her at dsenol@andrew.cmu.edu
\end{IEEEbiography}

\begin{IEEEbiography}{Jeremie Kim}{\,}is with ETH Zürich, Switzerland\aov{, and Carnegie Mellon University, USA}. Contact him at jeremie.kim@inf.ethz.ch
\end{IEEEbiography}

\begin{IEEEbiography}{Saugata Ghose}{\,}is with the University of Illinois at Urbana--Champaign and Carnegie Mellon University, USA. Contact him at ghose@illinois.edu
\end{IEEEbiography}

\begin{IEEEbiography}{Can Alkan}{\,}is with Bilkent University, Turkey. Contact him at calkan@cs.bilkent.edu.tr
\end{IEEEbiography}

\begin{IEEEbiography}{Onur Mutlu}{\,}is with ETH Zürich, Switzerland\ao{, Carnegie Mellon University, USA}\aol{, and Bilkent University, Turkey}. Contact him at \aov{omutlu@gmail.com}
\end{IEEEbiography}



\end{document}